\documentclass[aps,prl,twocolumn,showpacs,superscriptaddress]{revtex4-1}
\usepackage{bm}
\usepackage{mathrsfs}
\usepackage{amsmath}
\usepackage{amssymb}
\usepackage{graphicx}
\usepackage{amsfonts}
\usepackage{amsthm}
\usepackage{color}
\usepackage{dcolumn}
\usepackage{txfonts}

\begin{document}
\title{Solvable Lattice Gas Models With Three Phases}
\author{B. B. Wei}
\affiliation{Department of Physics, The Chinese University of Hong Kong, Hong Kong, China}
\author{C. N. Yang}
\email{Corresponding author: frankcnyang@gmail.com}
\affiliation{Department of Physics, The Chinese University of Hong Kong, Hong Kong, China}
\affiliation{Institute of Theoretical Physics, The Chinese University of Hong Kong, Hong Kong, China}
\affiliation{Institute for Advanced Study, Tsinghua University, Beijing, China}

\begin{abstract}
Phase boundaries in $p-T$ and $p-V$ diagrams are essential in material science researches. Exact analytic knowledge about such phase boundaries are known so far only in two-dimensional (2D) Ising-like models, and only for cases with two phases. In the present paper we present several lattice gas models, some with three phases. The phase boundaries are either analytically calculated or exactly evaluated.
\end{abstract}


\maketitle

In the 19-th century Maxwell's construction of vapor-liquid transition in the $p-V$ diagram of a gas was very famous. With the development of statistical mechanics it became possible in
20-th century to theoretically study such phase transitions. Utilizing the brilliant solution by Onsager\cite{Onsager1944} and Kaufman\cite{Kaufman1949} of the two-dimensional (2D) Ising model, a lattice gas model was constructed \cite{Lee1952b} in 1952 for which the two phase region in its $p-V$ diagram is \emph{analytically known}. In the present paper we construct duplex models which have three phases, not just two, and for which the phase boundary in $p-T$ diagrams and in $p-V$ diagrams can both be exactly calculated.

These models have two sublattices, and long range order form separately in the sublattices, creating a kind of \emph{partial order}.

\section{Models $A$ and $B$}
We shall refer to the model of paper II \cite{Lee1952b} for a square 2D lattice gas as model $A$. We shall adopt its notations, and refer to its equation (XX) as (II XX). The unit-circle theorem proved for that model guarantees that there can only be \emph{one phase transition between two phases}.  To go beyond that we now define a model for which the roots of its partition function lie on a circle of radius 1/2, not 1.

Consider a square 2D lattice, to be called model $B$, for which each site is occupied/vacant in three ways: vacant, or occupied in mode $\mu$ or occupied in mode $\nu$ (Notice it is
\emph{never} doubly occupied.)  Assume nearest neighbour atom-atom interaction with energy $-2\epsilon$, just like in model $A$.

We shall denote the grand partition function of these two models by $G_A(\epsilon;z,T)$ and $G_B(\epsilon;z,T)$. [$G_A$ was denoted by $\mathscr{P}$ in (II26).] Obviously,
\begin{eqnarray}\label{grandpartition}
G_B(\epsilon;z,T)=G_A(\epsilon;2z,T).
\end{eqnarray}
Thus for the case of $\epsilon>0$, all roots of $G_B$ lie on a circle centered at the origin in the complex $z$ plane with radius 1/2. For convenience we shall write
\begin{eqnarray}
Q(\epsilon;z,T)=\lim_{V\rightarrow\infty}\frac{1}{V}\ln G(\epsilon;z,T)
\end{eqnarray}
where $V$, the volume, is the total number of sites (called $\mathfrak{N}$ in Ref.\cite{Lee1952b}).  From Eq.\eqref{grandpartition} we have
\begin{eqnarray}\label{free energy}
Q_B(\epsilon;z,T)=Q_A(\epsilon;2z,T).
\end{eqnarray}
The grand partition function is related to the pressure $p$, the density $\rho$ and other thermodynamic variables by:
\begin{eqnarray}
\begin{array}{l}
\frac{p}{k_BT}=Q(\epsilon;z,T),\\
\rho=\frac{N}{V}=z\frac{\partial}{\partial z}Q(\epsilon;z,T),\\
\frac{E}{V}=k_BT^2\frac{\partial}{\partial T}Q(\epsilon;z,T).
\end{array}
\end{eqnarray}

\begin{figure}
\begin{center}
\includegraphics[scale=0.26]{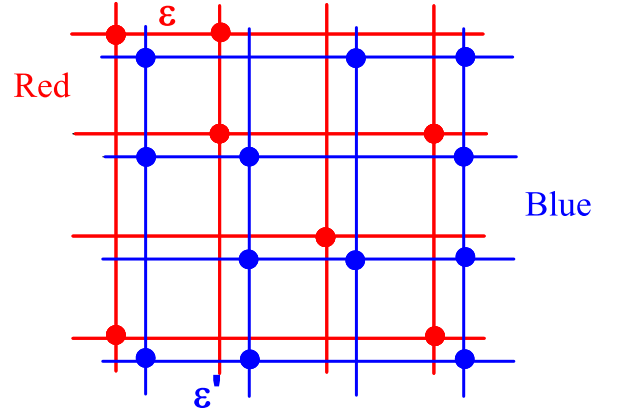}
\end{center}
\caption{(color online). Duplex model. The two sublattices have fixed ratio, $\alpha:\alpha'$, for their numbers of sites.}
\label{fig:epsart1}
\end{figure}

\section{Duplex Models}
A duplex model is one in which two lattice gas models are superposed on each other (Figure~\ref{fig:epsart1}). It can be a $A-A$, a $A-B$ or a $B-B$ combination. \emph{Interatomic} interaction \emph{within} each model, $A$ or $B$, is as described above. \emph{We assume no interatomic interactions} between the two constituent models. That assures that the grand partition function of the duplex is a \emph{product} of the grand partition functions of the two constituent models.

The number of sites in the two models will be kept at a constant ratio $\alpha:\alpha'$ (where $\alpha+\alpha'=1$) as both go to infinity. We designate such an $A-B$ duplex with coupling
constants $\epsilon_1$ and $\epsilon_2$ by
\begin{equation}\label{duplexmodel}
\{\alpha A(\epsilon_1),\alpha' B(\epsilon_2)\}
\end{equation}
Because of the product nature of its grand partition function, duplex model \eqref{duplexmodel} has as its $Q$ function a sum:
\begin{subequations}
\begin{eqnarray}
Q_D(\epsilon;z,T)&=&\alpha Q_A(\epsilon_1;z,T)+\alpha'Q_B(\epsilon_2;z,T)\nonumber\\
&=&\alpha Q_A(\epsilon_1;z,T)+\alpha'Q_A(\epsilon_2;2z,T)
\end{eqnarray}
Thus the thermodynamics of the duplex model \eqref{duplexmodel} can be evaluated from $Q_A(\epsilon;z,T)$.

We can also consider an $A-A$ duplex:
\begin{equation}\label{duplexmodel1}
\{\alpha A(\epsilon),\alpha' A(\epsilon')\}.\nonumber
\end{equation}
For such duplex model we have
\begin{eqnarray}
Q_D(\epsilon;z,T)&=&\alpha Q_A(\epsilon;z,T)+\alpha'Q_A(\epsilon';z,T). (6b)
\end{eqnarray}
\end{subequations}

\begin{figure}
\begin{center}
\includegraphics[width=\columnwidth]{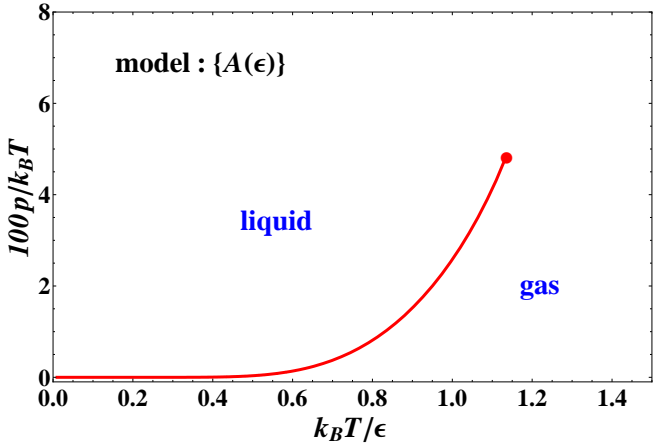}
\end{center}
\caption{(color online). $p-T$ diagram for $\{A(\epsilon)\}$ and $\{B(\epsilon)\}$. The critical point is given by equation (10).}
\label{fig:epsart2}
\end{figure}

\section{properties of $Q_A$}
(i) In 1941 Kramers and Wannier\cite{Kramers1941} discovered an important \emph{dual relationship} for the Ising model. For our model $A$ the relationship states that the $Q_A$ function at two temperatures, $T$ and $T^*$, are related:
\begin{eqnarray}
Q_A(\epsilon;1,T^*)=Q_A(\epsilon;1,T)+\ln[(1-x^2)/2x],
\end{eqnarray}
where
\begin{eqnarray}
\sinh\Big(-\frac{\epsilon}{k_BT}\Big)\sinh\Big(-\frac{\epsilon}{k_BT^*}\Big)=1,
\end{eqnarray}
and
\begin{eqnarray}
x=\exp\Big(-\frac{\epsilon}{k_BT}\Big).
\end{eqnarray}
The temperature $T$ and $T^*$ become identical when $x$ is equal to
\begin{eqnarray}
x_c=\sqrt{2}-1=\exp\Big(-\frac{\epsilon}{k_BT_c}\Big).
\end{eqnarray}

(ii) For model $A$, the grand partition function is a polynomial in $z$ where the coefficients of $z^a$ and $z^{V-a}$ are identical. Thus
\begin{eqnarray}
Q_A(\epsilon;z,T)-\frac{1}{2}\ln z=Q_A\Big(\epsilon;\frac{1}{z},T\Big)+\frac{1}{2}\ln z.
\end{eqnarray}
Taking the derivative with respect to $\ln z$ yields
\begin{eqnarray}\label{density}
\rho_A(\epsilon;z,T)=-\rho_A\Big(\epsilon;\frac{1}{z},T\Big)+1.
\end{eqnarray}
It was shown in paper II \cite{Lee1952b} that \eqref{density} leads to
\begin{eqnarray}
T\geq T_c, \ \ \rho_A(\epsilon;1,T)=1/2;
\end{eqnarray}
But for $T<T_c$, $\rho_A(\epsilon;z,T)$ is discontinuous at $z=1$, and \eqref{density} becomes
\begin{eqnarray}
T<T_c, \ \ \rho_A(\epsilon;1-,T)+\rho_A(\epsilon;1+,T)=1.
\end{eqnarray}
Furthermore, the value of $\rho_A(\epsilon;1-,T)$ is given explicitly as $v_g^{-1}$ in (II 15).

(iii) Combing the results of references \cite{Onsager1944,Kaufman1949, Lee1952b}, we know that function $Q_A(\epsilon;z,T)$, for physically relevant values of $\epsilon,z,$ and $T$,
is \emph{analytic everywhere} except on the half line $z=1$, and $0<x\leq \sqrt{2}-1 $. The value of $Q_A$ on this half line is explicitly known from \cite{Onsager1944,Kaufman1949}.
Its derivative with respect to $z$ is discontinuous on this half line, with the discontinuity explicitly evaluated in \cite{Lee1952b}.

Analytic evaluation of $Q_A(\epsilon;z,T)$ at any point where $z\neq1$ is at present not possible. Accurate numerical evaluation for $z\leq1/2$ can however be obtained from expansion
(II 18) which was originally due to Mayer\cite{Mayer1937}. Combing (II 23) and (II A) we have
\begin{eqnarray*}
y=x^4z.
\end{eqnarray*}
(II 18) becomes, for $0<z\leq1/2$,
\begin{eqnarray}
Q_A(\epsilon;z,T)&=&x^4z+\Big[2x^6-\frac{5}{2}x^8\Big]z^2+\Big[6x^8-16x^{10}+\frac{31}{3}x^{12}\Big]z^3\nonumber\\
& &+[x^8+\cdots]z^4+\cdots
\end{eqnarray}

\section{Phase Boundary Diagrams}
(i) The phase boundary in a $p-V$ diagram for model $A$ was explicitly given in paper II \cite{note}. The corresponding $p-T$ diagram is easily constructed from (II 14) and is given in Figure~\ref{fig:epsart2} above.
The phase boundary curve in Figure~\ref{fig:epsart2} is given by the equation
\begin{eqnarray}
p/k_BT=Q_A(\epsilon;1,T),
\end{eqnarray}
for $T$ values such that
\begin{eqnarray}
x=\exp(-\epsilon/k_BT)\leq x_c=\sqrt{2}-1.
\end{eqnarray}
The curve actually has a further portion for larger value of $x$. But this portion is not a phase boundary.  Thus it is not plotted.

\begin{figure}
\begin{center}
\includegraphics[width=\columnwidth]{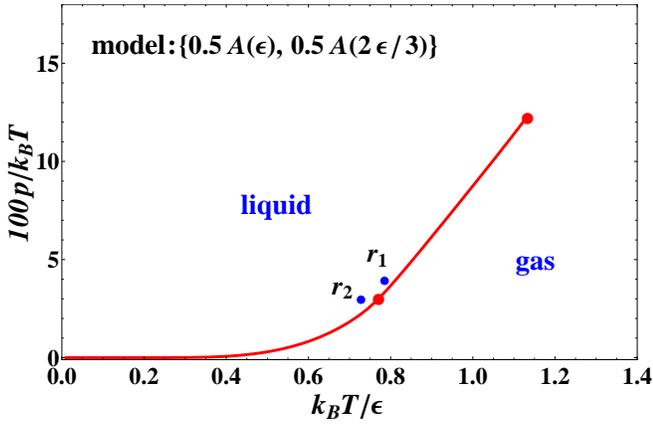}
\end{center}
\caption{(color online). $p-T$ diagram for an $A-A$ duplex. It has two critical points. The specific heat $C_V$ along the curve at constant $V/N=1/2$
 has logarithmic singularities at these two critical points. The points $r_1$ and $r_2$ have different long range orders. Cf. last section of the text.}
\label{fig:epsart3}
\end{figure}

\begin{figure}
\begin{center}
\includegraphics[width=\columnwidth]{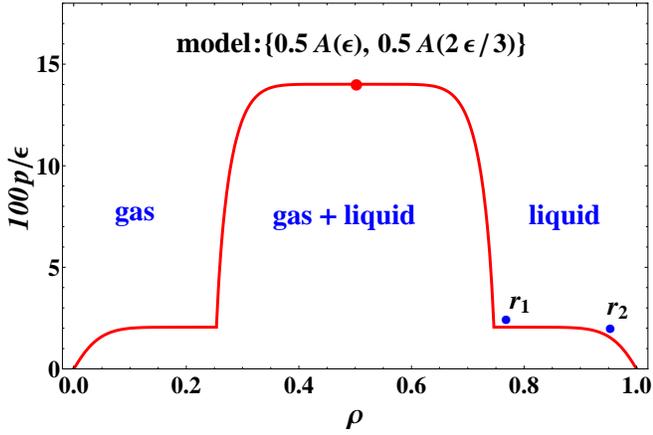}
\end{center}
\caption{(color online). $p-\rho$ diagram for an $A-A$ duplex.}
\label{fig:epsart4}
\end{figure}

For model $B$, Eq.\eqref{free energy} above shows that \emph{both $p-T$ and $p-V$ diagrams are identical to those for model $A$}.

(ii) Next we consider the duplex model $\{0.5A(\epsilon),0.5A(2\epsilon/3)\}$. With the $2\epsilon$'s we have now two critical temperatures $T_1$ and $T_2$, with $T_2=2T_1/3$.
The phase boundary is given by equation (6b) above at $z=1$:
\begin{eqnarray}
p/k_BT=0.5Q_A(\epsilon;1,T)+0.5Q_A(2\epsilon/3;1,T),
\end{eqnarray}
for $T\leq T_1$. This phase boundary is plotted in Figure~\ref{fig:epsart3}. Notice it has two singular points, at $T_2$ as well as at $T_1$.

The corresponding $p-\rho$ diagram is given in Figure~\ref{fig:epsart4}. Notice the \emph{symmetry} of the curve with respect to the vertical line $\rho=1/2$.

\begin{figure}
\begin{center}
\includegraphics[width=\columnwidth]{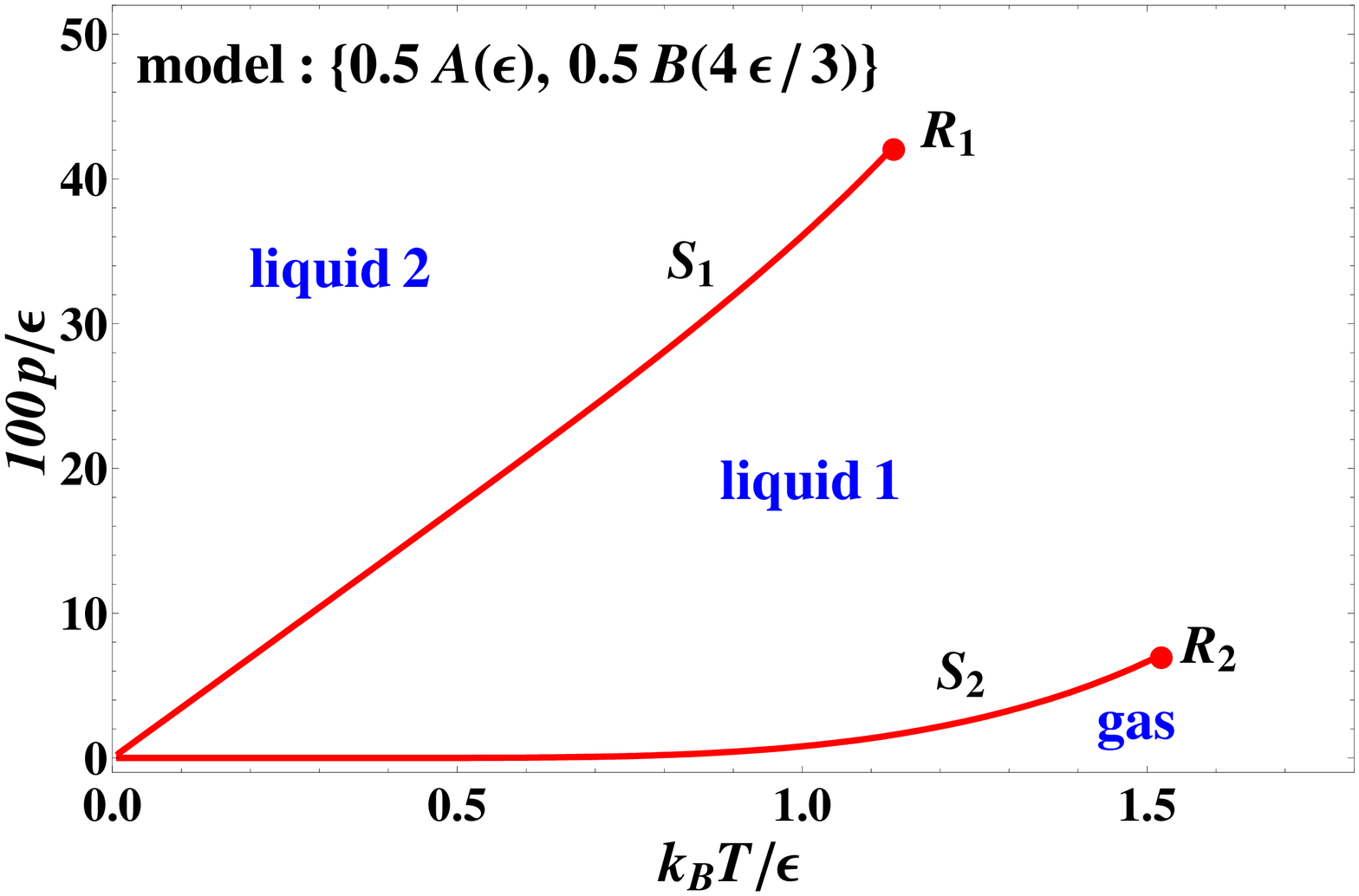}
\end{center}
\caption{(color online). $p-T$ diagram for an $A-B$ duplex. The line segments $\beta_1$ and $\beta_2$ of Fig.6 are respectively the lower and upper edges of
line $S_2$ in Fig.5. Similarly the line segments $\beta_3$ and $\beta_4$ of Fig.6 are respectively the lower and upper edges of line
$S_1$ in Fig.5.}
\label{fig:epsart5}
\end{figure}

(iii) This symmetry does not obtain for an $A-B$ duplex such as
\begin{eqnarray}\label{ABduplex}
\{0.5A(\epsilon),0.5B(4\epsilon/3)\}.
\end{eqnarray}
In this case there are again two critical temperatures $T_1$ and $T_2$ for the two sublattices, with $T_2=4T_1/3$. The $Q$ function of the duplex is given by (6a):
\begin{eqnarray}
p/k_BT=Q_D(z)=0.5Q_A(\epsilon;z,T)+0.5Q_A(4\epsilon/3;2z,T).
\end{eqnarray}
This equation shows that the duplex has two phase transitions. To see this, consider at a fixed $T$, $Q_D(z)$ as a function of $z$. It has singularities at $z=1$ and $2z=1$ for sufficiently low temperatures. These singular points are where phase transitions take place. The $z=1$ singularity in Figure~\ref{fig:epsart5} will be called line $S_1$, and the $z=1/2$ singularity will be called line $S_2$:
\begin{subequations}
Line $S_1$:
\begin{eqnarray}
 p/k_BT&=&0.5Q_A(\epsilon;1,T)+0.5Q_A(4\epsilon/3;2,T), [ie. z=1]\nonumber\\
\end{eqnarray}
Line $S_2$:
\begin{eqnarray}
 p/k_BT&=&0.5Q_A(\epsilon;1/2,T)+0.5Q_A(4\epsilon/3;1,T). [ie. 2z=1]\nonumber\\
\end{eqnarray}
\end{subequations}

\begin{figure}
\begin{center}
\includegraphics[width=\columnwidth]{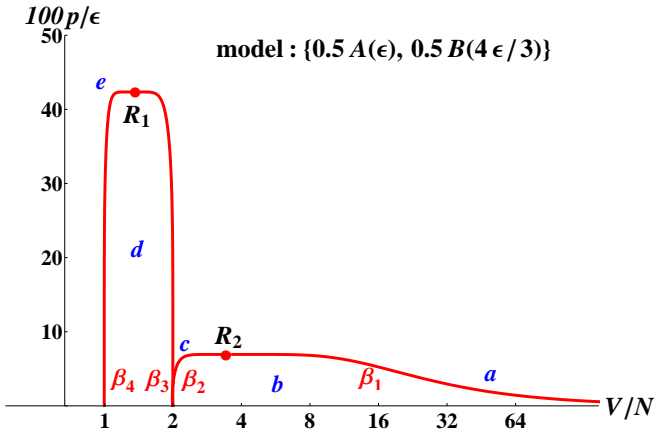}
\end{center}
\caption{(color online). $p-V$ diagram for an $A-B$ duplex. The five regions are: $a$=gas, $b$=gas+liquid 1,
$c$=liquid 1, $d$=liquid 1+liquid 2, and $e$=liquid 2.}
\label{fig:epsart6}
\end{figure}

Next we try to construct Figure~\ref{fig:epsart6}, the $p-V$ diagram for duplex model \eqref{ABduplex}. We first consider an isotherm at a low temperature $T<T_1<T_2$. According to Figure~\ref{fig:epsart5}, as we increase
the pressure starting from a low value, the duplex will undergo two phase transitions, first in crossing line $S_2$, then in crossing line $S_1$. The pressure at the first crossing is given
by (21b). The singularity at this crossing resides in the second term. The corresponding density jump from $z=1-$ to $z=1+$ is from
\begin{eqnarray}
\rho&=&0.5\rho_A(\epsilon;1/2,T)+0.5\rho_A(4\epsilon/3;1-,T),
\end{eqnarray}
to
\begin{eqnarray}
\rho&=&0.5\rho_A(\epsilon;1/2,T)+0.5\rho_A(4\epsilon/3;1+,T),
\end{eqnarray}
We thus have the pair of equations that define phase boundary $\beta_1$ in Figure~\ref{fig:epsart6}: (21b) and (22). Also for phase boundary $\beta_2$: (21b) and (23).

Similar reasonings yield the equations that define phase boundary $\beta_3$ in Figure~\ref{fig:epsart6} as the isotherm crosses line $S_1$. Boundary $\beta_3$:
\begin{equation}
\left\{
\begin{array}{c}
p/k_BT=0.5Q_A(\epsilon;1,T)+0.5Q_A(4\epsilon/3;2,T),\\
\rho=0.5\rho_A(\epsilon;1-,T)+0.5\rho_A(4\epsilon/3;2,T).
\end{array}
\right.
\end{equation}
For phase boundary $\beta_4$ we have\\Boundary $\beta_4$:
\begin{equation}
\left\{
\begin{array}{c}
p/k_BT=0.5Q_A(\epsilon;1,T)+0.5Q_A(4\epsilon/3;2,T),\\
\rho=0.5\rho_A(\epsilon;1+,T)+0.5\rho_A(4\epsilon/3;2,T).
\end{array}
\right.
\end{equation}

(iv) We now discuss some properties of the four phase boundaries $\beta_1,\beta_2,\beta_3,\beta_4$. (24) and (25) show that $\beta_3$ and $\beta_4$
would meet at $T_1$ where $\rho_A(\epsilon;1-,T_1)=\rho_A(\epsilon;1+,T_1)$. But both these $\rho$'s are equal to $1/2$ according to (13). Thus boundaries $\beta_3$ and $\beta_4$
meet at\\ $R_1$:
\begin{equation}
\left\{
\begin{array}{c}
p/k_BT=0.5Q_A(\epsilon;1,T_1)+0.5Q_A(4\epsilon/3;2,T_1), \\
\rho=0.5(1/2)+0.5\rho_A(4\epsilon/3;2,T_1),
\end{array}%
\right.
\end{equation}
Similarly boundaries $\beta_1$ and $\beta_2$ meet at \\$R_2$:
\begin{equation}
\left\{
\begin{array}{c}
p/k_BT=0.5Q_A(\epsilon;1/2,T_2)+0.5Q_A(4\epsilon/3;1,T_2),\\
\rho=0.5\rho_A(\epsilon;1/2,T_2)+0.5(1/2).
\end{array}
\right.
\end{equation}

(v) Referring to Figure~\ref{fig:epsart6}, a natural questions arises: Could line $\beta_2$ and $\beta_3$ intersect at a temperature $T_0>0$? The answer is no,
because if $T_0$ exists, then the $p$ value of line $\beta_2$, given by (21b) at that $T_0$ must be equal to that of line $\beta_3$, given by (24) at $T_0$. Because of
the monotonic property of $Q_A$ with respect to $z$, this is impossible.

But at $T=0$, they do intersect at point $R_3$ where $p/k_BT=0$, and $\rho=\alpha_2=0.5$.

(vi) In some cases, for example, for duplex model $\{0.8A(\epsilon),0.2B(4\epsilon/3)\}$, the $p-V$ diagram shows intersections of lines $\beta_2$ and $\beta_3$.
But such intersections take place at \emph{different $T$ }values on $\beta_2$ and $\beta_3$. Thus they disappear in the isotherm curves in a three dimensional $p-T-\rho$ diagram.

\section{Partial Order}
In Figure~\ref{fig:epsart5} and Figure~\ref{fig:epsart6} for an $A-B$ duplex there are two different liquid phases. Liquid 1 has long range order in the $B$ sublattices and no long range
order in $A$ sublattices. Liquid 2, on the other hand, has long range order in both sublattices. For the $A-A$ duplex there are boundary-less long rang order changes between points
$r_1$ and $r_2$ (Figure~\ref{fig:epsart3} and Figure~\ref{fig:epsart4}), In $r_1$ there is long range order in $\{A(\epsilon)\}$ but not in $\{A(2\epsilon/3)\}$, while in $r_2$ there is
long range order in both sublattices.

Thus the duplex structure allows for a special kind of partial order in which different sublattices exhibit different long range orders.

There has been many publications concerning phase transitions in lattice gas models, e.g. see \cite{partialorder}. The models in the present paper seem to be the few that deal with \emph{exactly
calculated} phase boundaries.

\section{Discussion}
(i) There are two key ideas in the present paper: Model $B$ and Duplex models. Are these ideas too far-fetched and not realizable? We think not: Model $B$ is in fact a special Potts model with $q=3$ \cite{Wu1982}. Our model $B$ is similar to a spin-1 system\cite{Griffiths1967,Wu1978}, the equivalence of model $B$ and model $A$ was first pointed by Griffiths\cite{Griffiths1967}. As to the duplex lattice illustrated in Figure~\ref{fig:epsart1}, one could construct a stratum consisting of two sublattices,
red and blue, and so that atoms on different sublattices do not interact.

(ii) Model $A$ is based on Onsager's solution of the Ising model. Long range correlation in the Ising model has been extensively studied in the 1960s\cite{Wu1973}, and provides
the basis for our understanding of gas-liquid phase transitions. To understand mathematically liquid-solid phase transitions, however, it is obvious that we need new models
which allow for solid-like long range correlations, going beyond all the models discussed in the present paper. How can we construct such new models? We speculate that adding
weak attractive next-to-nearest-neighbour interactions to model $A$, so as to favor the formation of small four atom squares, seems a good idea. It may be worthwhile to pursue computer studies of such models.

\begin{acknowledgements}
We are indebted to Professor R. B. Liu for discussions. This work is supported by the National Natural Science Foundation of China under Grants No. J1125001 and 11147002.
\end{acknowledgements}


\begin{references}
\bibitem{note}
The $p-V$ diagram in paper II is fine except for a typo: The label $p$ should read $(100)p$.
\end{references}
\end{document}